\documentclass[epj,nopacs]{svjour}

\usepackage{amsmath%,feynmp,slashed,subfig%
}
\usepackage[dvips]{graphicx}

\newcommand{\grav}{\mathcal{G}}
\newcommand{\kkm}{M_{\textrm{\tiny $K \! K$}}}
\newcommand{\cw}{\cos \theta_W}
\newcommand{\csw}{\cos^2 \theta_W}
\newcommand{\ssw}{\sin^2 \theta_W}

\newcommand{\smalltimes}%{\! \times \!}
{\times}

\begin{document}

\title{Production of a Z~boson and photon via a Randall-Sundrum-type graviton at
  the Large Hadron Collider}

\author{Jordan P.~Skittrall}

\institute{Department of
  Applied Mathematics and Theoretical Physics, Centre
  for Mathematical Sciences, University of Cambridge, Wilberforce Road, Cambridge, CB3~0WA,
  United Kingdom \email{J.P.Skittrall@damtp.cam.ac.uk}}

\abstract{In extra dimensional models with Kaluza-Klein graviton
  states that are well separated in mass, such states may be observed
  as resonances in collider experiments. We extend previous works on
  such scenarios by considering the one-loop resonant production of a
  Z~boson in association with a photon. We find the production rate to
be negligible in conservative scenarios, and small for
reasonable
luminosity in less conservative scenarios.
\keywords{Beyond Standard Model, Phenomenology of Field Theories in Higher Dimensions}}

%\preprint{DAMTP-2008-87}

\PACS{ {04.50.+h}{} \and {11.25.Mj}{} \and {12.15.-y}{} \and
  {12.38.Qk}{}}

\titlerunning{Production of a Z~boson and photon via a RS-type graviton at
  the LHC}
\authorrunning{Jordan P.\ Skittrall}

\maketitle

\section{Introduction}

A number of models, most notably the Randall-Sundrum~1 class
models~\cite{RaS1999},
predict the possibility of discrete TeV-scale graviton
resonances. The coupling of the Kaluza-Klein graviton states to matter is
enhanced when compared with the coupling of the ground state graviton
to matter~\cite{DaHR1999}. As a result, phenomenologically reasonable regions of
parameter space exist in which resonant tree-level production of
Standard Model particles via a Kaluza-Klein graviton excitation is
observable~\cite{AlOPW2000,AlOPPSW2002}. In a previous
work~\cite{AlSS2007} we extended an argument of Nieves and
Pal~\cite{NiP2005} to calculate the amplitude for the one-loop decay
of a
Z~boson to a photon and a Kaluza-Klein graviton, in a form applicable
to an effectively continuous spectrum of graviton excitations~\cite{AlS2007} in
the ADD model~\cite{ArDD1998,An1990,ArB1993,ArBQ1994}. With minimal
additional calculation it is possible to derive from this an amplitude
that can be used in calculations pertaining to the process $pp \to
\grav \to Z\gamma$, where $\grav$ is the first Kaluza-Klein excitation
of a graviton in a Randall-Sundrum model.

It will turn out that the one-loop suppression of this process is
sufficient to render the process negligible for most accessible
regions of parameter space. Further investigation of this process for
purposes of discovery or parameter constraint is
therefore not warranted, although observation of the process at high
luminosity would act as a consistency check on any claim of
observation of a Randall-Sundrum-type scenario. Observation of the
process at very low luminosity could not be explained by such a scenario.

This note is organised as follows. We first outline the theoretical
adjustments to our previous work required in order to calculate
\mbox{$\Gamma (\grav \to Z \gamma )$}. We then describe results
obtained for the cross-section \mbox{$\sigma ( pp \to \grav \to
  Z\gamma )$}, both analytically and using a modified version of
HERWIG~\cite{special:HERWIG6,special:HERWIG6.5}. We make a brief mention of the angular distribution of the process.

\section{Amplitude for the process $\grav \to Z \gamma$}

The amplitude for $\grav \to Z \gamma$ is closely related to the
amplitude for the process $Z \to \gamma \grav$ calculated in
reference~\cite{AlSS2007}, with the following choices/modifications:
\begin{itemize}
\item We choose $n=1$ extra dimensions, and consider only a single
  spin-2 Kaluza-Klein graviton excitation (i.e.\ we do not sum over
  excited states);
\item We set the gravitational coupling $\kappa$ equal to
  $1/\Lambda_{\pi}$, where
  $\Lambda_{\pi}=\kkm /((k/\bar{M}_{Pl})x_1)$, $x_1$ is the first
  root of the Bessel equation of order 1, $\kkm$ is the mass of the
  first Kaluza-Klein excitation of the graviton, and $k/\bar{M}_{Pl}$
  is the ratio of the Randall-Sundrum warp factor to the reduced
  Planck mass, for consistency
  with~\cite{AlOPW2000} chosen to satisfy $k/\bar{M}_{Pl}=0.01$ unless stated
  otherwise (this represents the value satisfying conservative
  theoretical constraints \cite{DaHR1999} that is most likely to give rise to
  detectable phenomena);
\item In this resonant production $\kkm \geq M_Z$, and we need to review
  one approximation in the analytic calculation, which was based on
  the assumption $M_Z \geq \kkm$;
\item The polarisation averaging is over the five polarisation states
  of the massive graviton, rather than the three of the Z~boson.
\end{itemize}

\subsection{The analytic approximation}

Reference~\cite{AlSS2007} gives the amplitude for the $\grav Z
\gamma$ interaction in the form
\begin{equation}
\mathcal{M}(q,k) = \mathcal{E}^{\lambda\rho}(q) \varepsilon^{\nu *}(k)
\varepsilon_Z^{\mu *}(p) F_{\lambda\rho\mu\nu}(q,k) \, ,
\end{equation}
with
$\mathcal{E}^{\lambda\rho}(q)$ and $q$ the polarisation tensor and
momentum respectively of the graviton, $\varepsilon^{\nu}(k)$ and $k$
the polarisation tensor and momentum respectively of the photon, and
$\varepsilon_Z^{\mu}(p)$ and $p$ the polarisation tensor and momentum
respectively of the Z~boson. From Ward-like identities, it can be
shown~\cite{AlSS2007} that
\begin{align}
F_{\lambda\rho\mu\nu} =& \left\{ (k_{\lambda}q_{\nu} -
k\cdot q \eta_{\nu\lambda}) (k_{\rho}q_{\mu} - k \cdot q
\eta_{\mu\rho} )F+ \right. \nonumber\\
& \phantom{\{} + \epsilon_{\lambda\nu\alpha\beta}q^{\alpha}k^{\beta} (k_{\rho}q_{\mu} -
k\cdot q \eta_{\mu\rho}) F_1+ \nonumber\\
& \phantom{\{}\left. +   (k_{\lambda}q_{\nu} - k\cdot q
\eta_{\nu\lambda}) \epsilon_{\rho\mu\alpha\beta}q^{\alpha}k^{\beta}F_2
\right\} + (\lambda \leftrightarrow \rho) \, \label{eq:generalform},
\end{align}
in which
\begin{align}
F_{\phantom{0}} =& \frac{\kappa e g}{4\pi^2 \cw} \times \nonumber\\
&\phantom{=} \times \left[ 
  \csw \left( 6 - \frac{1}{\csw} \right) J(M_W,\kkm,M_Z) - \phantom{\sum_f}\right.\nonumber\\
&\phantom{=\times [} \left. - 2 \sum_f Q_f X_f J(m_f,\kkm,M_Z) \right] \label{eq:bigfexpression}
,\\
F_1 =& 0 \, \label{eq:f1expression},\\
F_2 =& 0 \, \label{eq:f2expression},
\end{align}
and
\begin{equation}
J(X,Y,Z) = \int_0^1 \!\! dx \int_0^{1-x} \!\!\!\!\! dy \frac{x^2y(1-x-y)}{X^2 -
  y(1-x-y)Y^2 - xyZ^2} \,\label{eq:nastyintegral} .
\end{equation}
The integral of equation~\eqref{eq:nastyintegral} is not analytically
tractable, but can be solved by means of an approximation. The
approximate solutions derived in reference~\cite{AlSS2007} rely upon the assumption
$M_Z \geq \kkm$ (i.e.\ $Z \geq Y$), but it is not hard to replace this assumption with
the assumption $\kkm \geq M_Z$. To make the integral analytically
tractable, we need either to neglect the factor $X$ in the denominator
of the integrand (equivalent to assuming $X/(Y+Z) \ll
1/4$), or to neglect the factors $Y$ and $Z$ in that denominator
(equivalent to assuming $X/(Y+Z) \gg 1/4$). For each
integral in equation~\eqref{eq:bigfexpression} the most appropriate choice is to neglect the
factor $X$, but we note that this produces the constraint that the
result is valid only in the region $\kkm \gg 4 m_t$.
%for which it
%is sufficient to consider for all integrals we require the case $X/Y
%\ll 1/4$ ($\Rightarrow X/(Y+Z) \ll 1/4$).

%In this case, we may approximate the 
We therefore approximate each integral \linebreak[4] $J(M_X,\kkm,M_Z)$ by
$J(0,\kkm,M_Z)$ (where $M_X=M_W$ or $m_f$), and as in
reference~\cite{AlSS2007} this integral may be evaluated to obtain
\begin{align}
J(0,\kkm,M_Z) =& \frac{1}{12(M_Z^2-\kkm^2)} -
\frac{M_Z^2}{8(M_Z^2-\kkm^2)^2} + \nonumber\\ & +
\frac{\kkm^2M_Z^2}{4(M_Z^2-\kkm^2)^3} + \nonumber\\
 & + \frac{\kkm^4M_Z^2}{4(M_Z^2-\kkm^2)^4} \log \left(
\frac{\kkm^2}{M_Z^2} \right) .
\end{align}
We must perform a different series expansion of the logarithm from
that in reference~\cite{AlSS2007} in order to get a convergent series,
and the appropriate result is
\begin{align}
%\begin{equation}
J(0,\kkm,M_Z) = -\frac{1}{4} \sum_{j=0}^{\infty}& \frac{1}{(j+3)(j+4)}
\times \nonumber\\
& \times \kkm^{-2j-2}(\kkm^2-M_Z^2)^j \, .
%\end{equation}
\end{align}
This result agrees with the expansion obtained in
reference~\cite{AlSS2007} in the case $\kkm =M_Z$.

Returning now to equation~\eqref{eq:bigfexpression} we can substitute to obtain
\begin{align}
F =& \frac{\kappa e g}{4 \pi^2 \cw} \times \nonumber\\ &\times \left[ \csw \left(6 -
  \frac{1}{\csw} \right) -12 + 32 \ssw \right] \times \nonumber\\
 & \times \left[ - \frac{1}{4
    \kkm^2} \sum_{j=0}^{\infty} \frac{1}{(j+3)(j+4)}\left( 1 -
  \frac{M_Z^2}{\kkm^2} \right)^j \right] .
\end{align}
This expression is valid for $\kkm \gg 4 m_t$, and should be used with
caution for smaller graviton excitation masses.

Using the polarisation sum formulae quoted in
reference~\cite{AlSS2007}, we obtain
\begin{equation}
| \mathcal{M} |^2 = |F|^2 \left( \frac{(M_Z^2 - \kkm^2)^4 (7M_Z^2 +
  3\kkm^2)}{60 M_Z^2} \right) ,
\end{equation}
and hence
\begin{align}
%\begin{equation}
\Gamma(\grav \to Z \gamma) = \frac{1}{960 \pi M_Z^2 \kkm^3} & (\kkm^2 -
M_Z^2)^5 \times \nonumber\\ \times & (7M_Z^2 + 3\kkm^2) |F|^2 \, .
%\end{equation}
\end{align}

\section{The process $pp \to \grav \to Z \gamma$}

In order to investigate the process $pp \to \grav \to Z \gamma$, we
may follow one of two procedures:
\begin{enumerate}
\item Use a numerical calculation of $\sigma(pp \to \grav)$ and a \linebreak[4]
  numerical calculation of $\Gamma(\grav)$ in order to derive a
  \linebreak[4] cross-section for a given $\kkm$ via \\ \mbox{$\sigma(pp \to \grav \to Z
  \gamma) =$} \mbox{$\sigma(pp \to \grav) \Gamma(\grav \to Z \gamma)/
  \Gamma(\grav)$};
\item Use the expression for the amplitude for the process $\grav \to
  Z \gamma$ to make a narrow width approximation for the overall
  process, which may be used either directly for angular distribution
  calculations or via a numerical simulation to obtain values for
  $\sigma(pp \to \grav \to Z \gamma)$.
\end{enumerate}
We explore both possibilities. To perform the algebra necessary to
obtain the amplitude for the narrow width approximation, we use
FORM~\cite{special:FORM}, and for the numerical simulation, we use
HERWIG~6.5~\cite{special:HERWIG6,special:HERWIG6.5}, modified to be
able to calculate the cross-section for resonant $Z\gamma$
production. (The width for the decay $\grav \to Z \gamma$ is assumed
to be negligible in comparison with the overall width $\Gamma(\grav)$
and the implementation does not take into account the contribution of
$\Gamma(\grav \to Z \gamma)$ to $\Gamma(\grav)$, instead using the
HERWIG default of taking into account the contributions from
tree-level decays to partons, leptons and bosons.)

\begin{table}
\begin{center}
%\begin{scriptsize}
\begin{tabular}{| r @{.} l @{}| r @{.} l | r @{.} l | r @{.} l | r @{.} l %| r @{.} l
  |}
\hline
\multicolumn{2}%{@{}|@{}c@{}|@{}}
{|c|}%
{$\kkm$} & \multicolumn{2}{|c|}{$\sigma_{pp \to
  \grav}$} & \multicolumn{2}{|c|}{$\Gamma_{\grav}$} &
 \multicolumn{2}{|c|}{$BR_{\grav \to Z\gamma}$} &
 \multicolumn{2}{|c|}{$\sigma_{pp \to \grav \to Z \gamma}$} %&
% \multicolumn{2}{|c|}{$\sigma_{pp \to
%  \grav \to Z\gamma}$}
\\
%\cline{3-10}
\multicolumn{2}%{@{}|@{}c@{}|@{}}
{|c|}%
{(TeV)} & \multicolumn{2}{|c|}{(pb)} &
\multicolumn{2}{|c|}{{%\scriptsize
{(GeV)}}} &
\multicolumn{2}{|c|}{%(analytic)
} & \multicolumn{2}{|c|}{(pb)} %&
%\multicolumn{2}{|c|}{(pb)}
\\
%\multicolumn{2}{@{}|@{}c@{}|@{}}{} & \multicolumn{2}{|c|}{(comp)} &
%\multicolumn{2}{|c|}{{\scriptsize{(comp)}}} & \multicolumn{2}{|c|}{} & \multicolumn{2}{|c|}{(analytic)} &
%\multicolumn{2}{|c|}{(comp)}\\
\hline
$\mathbf{0}$&$\mathbf{3}$ & \multicolumn{2}{|l|}{$\mathbf{86}$} &
$\mathbf{0}$&$\mathbf{041}$ & $\mathbf{8}$&$\mathbf{4 \smalltimes
10^{-9}}$ & $\mathbf{7}$&$\mathbf{2 \smalltimes 10^{-7}}$ %&
%$\mathbf{7}$&$\mathbf{4 \smalltimes 10^{-7}}$
\\
$0$&$5$ & \multicolumn{2}{|l|}{$10$} & $0$&$068$ &
$4$&$1 \smalltimes 10^{-8}$ & $4$&$2 \smalltimes
10^{-7}$ %& $4$&$2 \smalltimes 10^{-7}$
\\
$0$&$9$ & $0$&$68$ & $0$&$127$ & $1$&$8 \smalltimes 10^{-7}$ & $1$&$2
\smalltimes 10^{-7}$ %& $1$&$2 \smalltimes 10^{-7}$
\\
$1$&$0$ & $\phantom{1}0$&$41$ & $0$&$141$ & $2$&$2 \smalltimes 10^{-7}$ &
$9$&$2 \smalltimes 10^{-8}$ %& $9$&$3 \smalltimes 10^{-8}$
\\
$1$&$5$ & $\phantom{1}0$&$05$ & $0$&$213$ & $5$&$5 \smalltimes 10^{-7}$ & $2$&$7
\smalltimes 10^{-8}$ %& $2$&$8 \smalltimes 10^{-8}$
\\
$1$&$7$ & $\phantom{1}0$&$025$ & $0$&$242$ & $7$&$1 \smalltimes 10^{-7}$ & $1$&$8
\smalltimes 10^{-8}$ %& $1$&$8 \smalltimes 10^{-8}$
\\
$1$&$8$ & $\phantom{1}0$&$018$ & $0$&$256$ & $8$&$1 \smalltimes 10^{-7}$ & $1$&$4
\smalltimes 10^{-8}$ %& $1$&$5 \smalltimes 10^{-8}$
\\
$1$&$9$ & $\phantom{1}0$&$013$ & $0$&$270$ & $9$&$1 \smalltimes 10^{-7}$ & $1$&$2
\smalltimes 10^{-8}$ %& $1$&$2 \smalltimes 10^{-8}$
\\
$2$&$0$ & $\phantom{1}0$&$0096$ & $0$&$285$ & $1$&$0 \smalltimes 10^{-6}$ & $9$&$7
\smalltimes 10^{-9}$ %& $9$&$9 \smalltimes 10^{-9}$
\\
$2$&$1$ & $\phantom{1}0$&$0071$ & $0$&$298$ & $1$&$1 \smalltimes 10^{-6}$ & $8$&$0
\smalltimes 10^{-9}$ %& $8$&$1 \smalltimes 10^{-9}$
\\
$2$&$2$ & $\phantom{1}0$&$0054$ & $0$&$312$ & $1$&$2 \smalltimes 10^{-6}$ & $6$&$6
\smalltimes 10^{-9}$ %& $6$&$8 \smalltimes 10^{-9}$
\\
\hline
\end{tabular}
%\end{scriptsize}
\caption{Masses $\kkm$ and cross-sections and widths of
  graviton resonances simulated with HERWIG %
  (columns 2--3)% (comp)
  , together with branching ratios and cross
  sections for decay to $Z\gamma$ calculated using the simulated total
  cross-section and width with an analytic calculation of the
  $Z\gamma$ decay width (columns 4--5; procedure~1 as outlined in the text)%, and with completely simulated
%  cross-sections
  . %
  The simulated values for the overall cross-section obtained using a
  narrow width approximation (procedure~2
  as outlined in the text) are substantially similar to those in the
  last column of this table, and are therefore omitted. The values in the $\Gamma_{\grav}$ column are taken
  from reference~\cite{AlOPW2000}. All values relate to 14~TeV $pp$
  collisions with $k/\bar{M}_{Pl}=0.01$. The row printed in bold type
  corresponds to the minimum value of $\kkm$ at 95\% level, according
  to reference~\cite{Abetal2007}. The results should be used
  with caution for values of $\kkm$ that are not significantly above
  $800$~GeV:
  see the text for discussion.}\label{table:cross_section_results}
\end{center}
\end{table}
Table~\ref{table:cross_section_results} contains results for the
benchmark value of $k/\bar{M}_{Pl}=0.01$, and
Table~\ref{table:cross_section_results_optimistic} contains the
results for the more experimentally accessible but less conservative
value of $k/\bar{M}_{Pl}=0.1$ (representing the largest feasible value
of $k/\bar{M}_{Pl}$~\cite{DaHR2000}). In each table the row
printed in bold corresponds to the lightest Kaluza-Klein graviton mass
not excluded by experiment~\cite{Abetal2007,Aaetal2007} -- and
therefore the point for
which it would be easiest to observe a deviation from the Standard
Model.

\begin{table}
\begin{center}
%\begin{scriptsize}
\begin{tabular}{| r @{.} l @{}| r @{.} l | r @{.} l | r @{.} l | r @{.} l %| r @{.} l
  |}
\hline
%\multicolumn{2}{|c|}{$\kkm$} & \multicolumn{2}{|c|}{$\sigma_{pp \to
%  \grav}$ (pb)} & \multicolumn{2}{|c|}{$\Gamma_{\grav}$ (GeV)} &
% \multicolumn{2}{|c|}{$BR_{\grav \to Z\gamma}$} &
% \multicolumn{2}{|c|}{$\sigma_{pp \to \grav \to Z \gamma}$
%    (pb)} & \multicolumn{2}{|c|}{$\sigma_{pp \to
%  \grav \to Z\gamma}$ (pb)}\\
%\cline{3-10}
%\multicolumn{2}{|c|}{(TeV)} & \multicolumn{2}{|c|}{(HERWIG)} &
%\multicolumn{2}{|c|}{(HERWIG)} & \multicolumn{2}{|c|}{(analytic)} & \multicolumn{2}{|c|}{(analytic)} &
%\multicolumn{2}{|c|}{(HERWIG)}\\
\multicolumn{2}%{@{}|@{}c@{}|@{}}
{|c|}%
{$\kkm$} & \multicolumn{2}{|c|}{$\sigma_{pp \to
  \grav}$} & \multicolumn{2}{|c|}{$\Gamma_{\grav}$} &
 \multicolumn{2}{|c|}{$BR_{\grav \to Z\gamma}$} &
 \multicolumn{2}{|c|}{$\sigma_{pp \to \grav \to Z \gamma}$} %&
% \multicolumn{2}{|c|}{$\sigma_{pp \to
%  \grav \to Z\gamma}$}
\\
%\cline{3-10}
\multicolumn{2}%{@{}|@{}c@{}|@{}}
{|c|}%
{(TeV)} & \multicolumn{2}{|c|}{(pb)} &
\multicolumn{2}{|c|}{{%\scriptsize
{(GeV)}}} &
\multicolumn{2}{|c|}{%(analytic)
} & \multicolumn{2}{|c|}{(pb)} %&
%\multicolumn{2}{|c|}{(pb)}
\\
%\multicolumn{2}{@{}|@{}c@{}|@{}}{} & \multicolumn{2}{|c|}{(comp)} &
%\multicolumn{2}{|c|}{{\scriptsize{(comp)}}} & \multicolumn{2}{|c|}{} & \multicolumn{2}{|c|}{(analytic)} &
%\multicolumn{2}{|c|}{(comp)}\\
\hline
$0$&$3$ & \multicolumn{2}{|l|}{$8200$} & $\phantom{0}4$&$1$ & $8$&$4 \smalltimes
10^{-9}$ & $6$&$9 \smalltimes 10^{-5}$ %& $7$&$1 \smalltimes 10^{-5}$
\\
$0$&$5$ & \multicolumn{2}{|l|}{$\phantom{0}970$} & $\phantom{0}6$&$9$ &
$4$&$1 \smalltimes 10^{-8}$ & $3$&$9 \smalltimes
10^{-5}$ %& $4$&$1 \smalltimes 10^{-5}$
\\
$\mathbf{0}$&$\mathbf{9}$ &
\multicolumn{2}{|l|}{$\mathbf{\phantom{00}65}$} &
\multicolumn{2}{|l|}{$\mathbf{13}$} & $\mathbf{1}$&$\mathbf{8 \smalltimes
  10^{-7}}$ & $\mathbf{1}$&$\mathbf{1
\smalltimes 10^{-5}}$ %& $\mathbf{1}$&$\mathbf{2 \smalltimes 10^{-5}}$
\\
$1$&$0$ & \multicolumn{2}{|l|}{$\phantom{00}39$} & \multicolumn{2}{|l|}{14} & $2$&$2 \smalltimes 10^{-7}$ &
$8$&$7 \smalltimes 10^{-6}$ %& $8$&$9 \smalltimes 10^{-6}$
\\
$1$&$5$ & $\phantom{000}4$&$8$ & \multicolumn{2}{|l|}{21} & $5$&$5 \smalltimes 10^{-7}$ & $2$&$6
\smalltimes 10^{-6}$ %& $2$&$7 \smalltimes 10^{-6}$
\\
$1$&$7$ & $\phantom{000}2$&$4$ & \multicolumn{2}{|l|}{24} & $7$&$2 \smalltimes 10^{-7}$ & $1$&$7
\smalltimes 10^{-6}$ %& $1$&$7 \smalltimes 10^{-6}$
\\
$1$&$8$ & $\phantom{000}1$&$7$ & \multicolumn{2}{|l|}{26} & $8$&$1 \smalltimes 10^{-7}$ & $1$&$4
\smalltimes 10^{-6}$ %& $1$&$4 \smalltimes 10^{-6}$
\\
$1$&$9$ & $\phantom{000}1$&$2$ &  \multicolumn{2}{|l|}{27} & $9$&$1 \smalltimes 10^{-7}$ & $1$&$1
\smalltimes 10^{-6}$ %& $1$&$2 \smalltimes 10^{-6}$
\\
$2$&$0$ & $\phantom{000}0$&$92$ &  \multicolumn{2}{|l|}{28} & $1$&$0 \smalltimes 10^{-6}$ & $9$&$3
\smalltimes 10^{-7}$ %& $9$&$4 \smalltimes 10^{-7}$
\\
$2$&$1$ & $\phantom{000}0$&$69$ &  \multicolumn{2}{|l|}{30} & $1$&$1 \smalltimes 10^{-6}$ & $7$&$7
\smalltimes 10^{-7}$ %& $7$&$8 \smalltimes 10^{-7}$
\\
$2$&$2$ & $\phantom{000}0$&$51$ &  \multicolumn{2}{|l|}{31} & $1$&$2 \smalltimes 10^{-6}$ & $6$&$3
\smalltimes 10^{-7}$ %& $6$&$4 \smalltimes 10^{-7}$
\\
\hline
\end{tabular}
%\end{scriptsize}
\caption{
%Masses $\kkm$ and cross-sections and widths of
%  graviton resonances simulated with HERWIG (comp), together with branching ratios and cross
%  sections for decay to $Z\gamma$ calculated using the simulated total
%  cross-section and width with an analytic calculation of the
%  $Z\gamma$ decay width, and with completely simulated
%  cross-sections.  All values relate to 14TeV $pp$
%  collisions with $k/\bar{M}_{Pl}=0.1$. The row printed in bold type
%  corresponds to the minimum value of $\kkm$ at 95\% level, according
%  to reference~\cite{Abetal2007}. The results should be used
%  with caution for values of $\kkm$ that are not significantly above
%  $800$GeV:
%  see the text for discussion.
Cross-sections, branching ratios and widths for the process $pp\to
\grav \to Z\gamma$ corresponding to
Table~\ref{table:cross_section_results}, but for the case  $k/\bar{M}_{Pl}=0.1$.
}\label{table:cross_section_results_optimistic}
\end{center}
\end{table}

For the entire range of
Kaluza-Klein masses considered in
Table~\ref{table:cross_section_results}, the cross-section is too small for a
non-negligible event rate at the LHC\@. For some of the mass range
considered in Table~\ref{table:cross_section_results_optimistic}, the
resultant event rates are not trivially negligible with $3000$~fb$^{-1}$ of data from a sLHC upgrade -- and it is reasonable from a
statistical perspective to consider the event rates of the individual
points, assuming that this channel would be used as a consistency
check following discovery in another channel.

The most favourable Kaluza-Klein
mass considered in Table~\ref{table:cross_section_results_optimistic}
that is not already excluded ($900$~GeV) yields an event rate too small for
observation at the LHC, but an event rate that with $3000$~fb$^{-1}$
of data from a sLHC upgrade needs slightly more careful consideration
($\sim 35$). However, the number of background events at this
integrated luminosity is relatively high in comparison with the
signal: a very rough estimate using
ALPGEN~\cite{special:ALPGEN}\footnote{The parameters used were:
  events with 1~photon, 1~Z~boson (all decay modes -- for comparison with the
  all-decays $pp \to \grav \to Z\gamma$ cross-section), and no
  additional hard jets, with
  a $p_T$ cut on the Z~boson of 40\% of the relevant graviton mass
  $\kkm$. This is merely a very rough indicative estimate of
  background -- but is sufficient to demonstrate that with a significant
  number of background events it would be very difficult to infer the
  existence of a signal.} gives an approximate background cross-section of
$4.0\times 10^{-5}$~pb ($\sim 121$ events with $3000$~fb$^{-1}$
of data), and when one additionally takes into
account detector effects and the necessity of only looking at a subset
of Z~decay channels 
for event tagging, one is led to the conclusion that it would be very
difficult to observe this process at the LHC, even with a luminosity upgrade.

\subsection{Angular distributions}
Although moot in many regions of accessible parameter space, the angular
distributions of the process may be relevant in particularly favourable
regions of parameter space. It is possible to calculate the angular
distributions algebraically~\cite{special:FORM} given the general form
of the $\grav Z \gamma$ vertex given in
equation~\eqref{eq:generalform}, together with
equations~\eqref{eq:f1expression} and~\eqref{eq:f2expression}, and
expressions for the $q\bar{q}\grav$ and $gg\grav$ vertices and
graviton propagator, obtainable from
references~\cite{GiRW1998,HaLZ1998}, for example.

\begin{table}
\begin{center}
%\begin{tabular}{|c|c|c|}
%\hline
%Process & Distribution & $\beta' = 1$ (high $\kkm$ limit)\\
%\hline
%$gg \to \grav \to Z\gamma$ & $2(1+\cos^2 \theta^* ) - \beta' (1 + \cos^2
%\theta^* )^2$ & $1-\cos^4 \theta^*$ \\
%$q\bar{q} \to \grav \to Z\gamma$ & $3 - \cos^2 \theta^* - 2\beta' (1 +
%\cos^2 \theta^* - 2 \cos^4 \theta^* )$ & $1 - 3\cos^2 \theta^* + 4
%\cos^4 \theta^*$\\
%\hline
%\end{tabular}
\begin{tabular}{|c|c|}
\hline
Process & Distribution\\
\hline
$gg \to \grav \to Z\gamma$ & $2(1+\cos^2 \theta^* ) - \beta' (1 + \cos^2
\theta^* )^2$\\
$q\bar{q} \to \grav \to Z\gamma$ & $3 - \cos^2 \theta^* - 2\beta' (1 +
\cos^2 \theta^* - 2 \cos^4 \theta^* )$\\
\hline
\end{tabular}
%\begin{tabular}{|c|c|c|}
%\hline
%Process & Distribution with $\beta' = 1$ (high $\kkm$ limit)\\
%\hline
%$gg \to \grav \to Z\gamma$ & $1-\cos^4 \theta^*$ \\
%$q\bar{q} \to \grav \to Z\gamma$ & $1 - 3\cos^2 \theta^* + 4
%\cos^4 \theta^*$\\
%\hline
%\end{tabular}
\caption{Angular distributions for the hard subprocesses of resonant
  $Z\gamma$ production via a Kaluza-Klein graviton. $\theta^*$ is the
  polar angle of the outgoing particle in the graviton rest frame;
  $\beta'$ is equal to $1 - M_Z^2/\kkm^2$ (so in most cases we can take
  $\beta' = 1$).}\label{tab:angulardistributions}
\end{center}
\end{table}

\begin{figure}
\begin{center}
\includegraphics[width=0.45\textwidth]{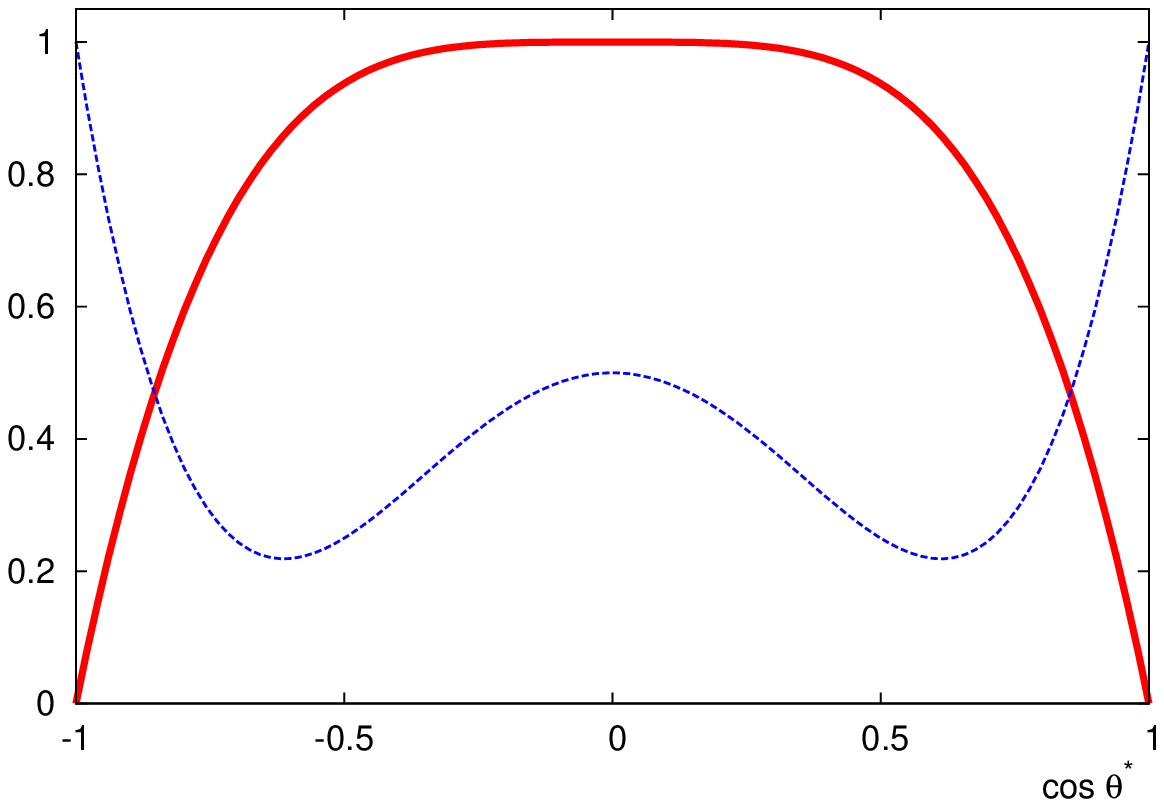}
\caption{Angular distributions for the subprocesses \mbox{$gg \to \grav \to
  Z\gamma$} (thick solid line) and \mbox{$q\bar{q} \to \grav \to Z\gamma$}
  (thin dashed line), in the limit $\kkm \gg M_Z$ ($\beta'=1$), normalised to have maximum value $1$ in each
case.}\label{fig:angulardistributions}
\end{center}
\end{figure}

The angular distributions are given in
Table~\ref{tab:angulardistributions} and are plotted for the
high-$\kkm$ limit in
Figure~\ref{fig:angulardistributions}.

Both $Z\gamma$ decay subprocesses differ in angular distribution from
the respective
$\gamma\gamma$ and $ZZ$ decay subprocesses (see
reference~\cite{AlOPPSW2002}).  This should not be surprising, given
that the $\gamma\gamma$ and $ZZ$ decay subprocesses differ in angular
distribution from each other, owing to the additional polarisation
states of the Z~bosons.

\section{Conclusions}
We have evaluated branching ratios and cross-sections for the hadronic
production of a Z~boson and a photon via a single Kaluza-Klein
graviton excitation, applicable to scenarios that lead to discrete
graviton resonances. We find that the process is unlikely to be
observed at the Large Hadron Collider, although its presence or
absence would be a potentially useful consistency check were other signals of
such models seen, and the process would provide additional data to
constrain the gravitational coupling $1/\Lambda_{\pi}$ (since this
coupling is dependent upon the Kaluza-Klein graviton mass and determines the
event rate, so the coupling is related to the energy scale and number of resonant events).

\begin{acknowledgement}
I should like to thank the members of the Cambridge Supersymmetry
Working Group, and in particular Ben Allanach, Are Raklev and Bryan Webber, for
helpful comments. I should also like to thank the organisers of the Summer
Institute 2008 in
Chi-Tou, Taiwan, for their hospitality whilst this paper was in
preparation. I am supported by the United Kingdom's Science
and Technology Facilities Council.
\end{acknowledgement}

\bibliographystyle{elsarticle-num}
\bibliography{abbrev_short,gtozphoton_bib}

\end{document}